\documentclass[preprint,showpacs,preprintnumbers,amsmath,amssymb]{revtex4}

\usepackage{graphicx}% Include figure files
\usepackage{dcolumn}% Align table columns on decimal point
\usepackage{bm}% bold math

\begin{document}

\title{Pushing the detection limit of Magnetic Circular Dichroism to 2 nm}

\author{Peter Schattschneider}
 \altaffiliation[Also at ]{Institute for Solid State Physics, Vienna University of Technology, Wiedner Hauptstrasse 8-10/138, A-1040 Vienna, Austria}
 
\author{Michael St{\"o}ger-Pollach}
  \affiliation{University Service Center for Transmission Electron Microscopy, Vienna University of Technology, Wiedner Hauptstrasse 8-10/052, A-1040 Vienna, Austria}%

\author{Stefano Rubino}
\email{stefanorubino@yahoo.it}
 \affiliation{Institute for Solid State Physics, Vienna University of Technology, Wiedner Hauptstrasse 8-10/138, A-1040 Vienna, Austria}%
 
 \author{Matthias Sperl, Christian Hurm, Josef Zweck}
  \affiliation{Institut f{\"u}r experimentelle und angewandte Physik, University of Regensburg,
D-93047 Regensburg, Germany}%

 \author{J\'{a}n Rusz}
  \affiliation{Department of Physics, Uppsala University, Box 530, S-751~21 Uppsala, Sweden}%

\date{\today}

\begin{abstract}
Magnetic Circular Dichroism (MCD) is a standard technique for the study of magnetic properties of materials in synchrotron beamlines. We present here a new scattering geometry in the Transmission Electron Microscope through which MCD can be observed with unprecedented spatial resolution. A convergent electron beam is used to scan a multilayer Fe/Au sample and record energy loss spectra. Differences in the spectra induced by the magnetic moments of the Fe atoms can be resolved with a resolution of 2 nm. This is a breakthrough achievement when compared both to the previous EMCD resolution (200 nm) or the best XMCD experiments (approx. 20 nm), with an improvement of two and one order of magnitude, respectively. \end{abstract}

\pacs{Valid PACS appear here}% PACS, the Physics and Astronomy
\maketitle

Detection of Magnetic Circular Dichroism (MCD) in the electron microscope was first reported in 2006~\cite{Nature2006}, providing an alternative to X-ray MCD (XMCD), the standard technique for investigation of spin and orbital magnetic moments in the synchrotron. The fact that spin-related properties can now be studied with a  commercial transmission electron microscope equipped with an energy spectrometer or an energy filter explains the  attraction of this novel technique, named Energy-loss Magnetic Chiral Dichroism (EMCD). 

Contrary to XMCD that measures polarization dependent X-ray absorption cross sections~\cite{SchutzPRL87,ChenPRB90,StohrS93,FischerJPD98}, EMCD exploits faint differences in the double differential scattering cross section (DDSCS) of fast electrons in the diffraction plane of an electron microscope. 

The main difficulty in EMCD is the low count rate which has limited the spatial resolution to  200~nm when the method was introduced a year ago~\cite{Nature2006}.
% using the spectrometer shift method. 
In the meantime, several labs have adopted EMCD~\cite{WarotUM07,CEMES,vanAkenMM07}, and it has become clear that at least two more scattering geometries can be used. The spatial resolution was improved to 30~nm with the LACDIF method~\cite{LACDIF}; here we report a substantial improvement that allows a spatial resolution of  2~nm to be attained by using convergent electron beam diffraction in the scanning mode of the TEM. 

The standard technique for the study of magnetic dichroism on a sub-micrometer scale is XMCD microscopy (typically with resolutions of about 25 to 50~nm), based on obtaining a circular dichroic signal in combination with imaging optics in  a synchrotron. This can be achieved either  with electron-optical lenses to form images with  photoemitted electrons (XMCD-PEEM) or with diffractive X-ray optics   ~\cite{StohrS93,SchneiderAPL93,StohrSRL98,TonnerRSI88,OhldagPRL01,ImadaSRL02} where a resolution of 15~nm has been reported~\cite{KimJAP06}. Also lensless imaging techniques appear to be very promising, in particular for time-resolved experiments~\cite{EisebittNat2004}.

These techniques have led to considerable progress in the understanding of magnetism in the solid state, and they become increasingly important for the rapidly expanding field of spintronics. The demand for extremely high spatial resolution that  arises in this context is met by the intrinsic sub-nm resolution of the TEM. 

Here we present a new geometry that allows the detection of  EMCD in the TEM on the nanometer scale. The method is applied to an epitaxial multilayer of Fe/Au, demonstrating a spatially resolved MCD signal on a 3 nm wide Fe layer. An analysis  of the signal-to-noise Ratio (SNR) shows that the spatial resolution for the detection of MCD is 2~nm with the chosen set-up.
% this value is likely to be better in a C$_S$ corrected microscope.

\begin{figure}
\centerline{\includegraphics[width=\textwidth]{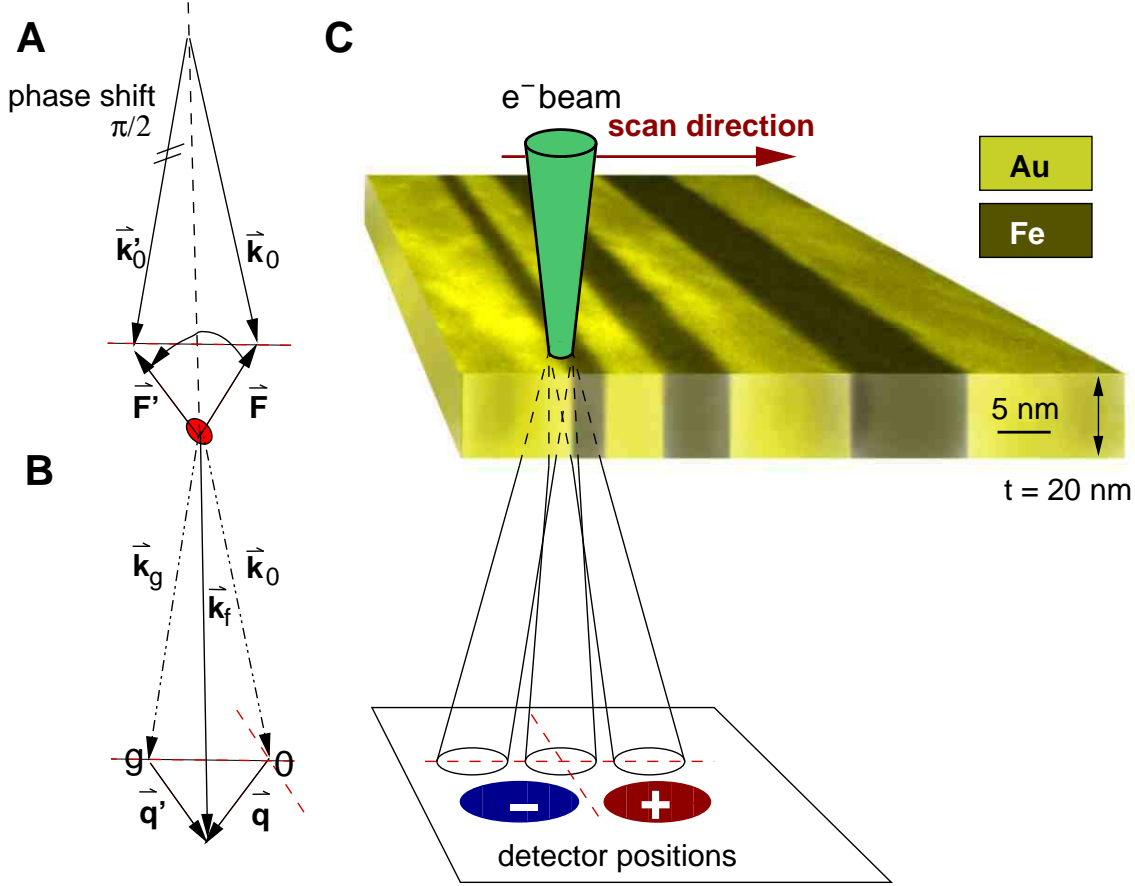}}
\caption{Principle of EMCD. \textbf{A:} Two coherent incident plane waves, dephased by $\pi/2$, produce a rotating electric perturbation $\vec F$ during the atomic excitation. \textbf{B:} When crystal diffraction is used, the detector position with respect to the 0 and g beam determines the final scattering direction $\vec k_f$ and thus $\vec q$ and $\vec q'$. \textbf{C:} A convergent electron beam is scanned across the Au/Fe multilayer sample. The detector is alternatively placed at positions "+" and "-" in the diffraction plane selecting two scattering vectors $q \perp q'$. The specimen image is a high-angle annular dark field (HAADF) map of the multilayer taken with the 1.7~nm electron probe.}
\label{fig1}
\end{figure}

The DDSCS in the dipole approximation for a geometry with two coherent incident plane waves $\vec k_0$, $\vec k_0'$  phase shifted by $\pm \pi/2$ (Fig. 1A) is~\cite{HebertUM03,SchattUM07}
\begin{equation}
\frac{\partial^2\sigma}{\partial E\partial\Omega}
=
\frac{4 \gamma^2}{a_0^2 }  \frac{k_f}{k_0}
\big(\frac{S({\vec q,\vec q },E)}{q^4} + \frac{S(\vec {q'},\vec{q'},E)}{q'\, ^4} \mp 2 \frac{\Im[S({\vec q }, \vec 
{q'},E)}{q^2q'^2}\big)
\label{DDSCSinter}
\end{equation}
where
\begin{equation}
S(\vec q , \vec {q '},E):=
\sum_{i,f}
 \langle f | \vec{q }\cdot\vec{R}| i \rangle
\langle i | \vec{q '}\cdot\vec{R}| f \rangle
\delta(E_i-E_f+E)
\label{MDFF}
\end{equation}
is the  mixed dynamic form factor~\cite{KohlUM85,SchattPRB99}, $\vec {k_f}$ is the final scattering wave vector (defined by the detector position), $\vec q = \vec {k_f}-\vec {k_0}$ and $\vec q'= \vec {k_f}-\vec {k'_0}$ are the wave vector transfers, $\gamma$ is the relativistic factor,  $a_0$ the Bohr radius and $E$ is the energy lost by the fast electron. Using the crystal lattice as a beam splitter the coherent incident waves ${\vec k}_0,  \, {\vec k'}_0 $ are replaced by ${\vec k}_0,  \, {\vec k}_g ={\vec k}_0 + {\vec g} $ with a reciprocal lattice vetor $\vec g$, as drawn in Fig. 1B.

For an atom at the origin the quantity $S$ possesses an imaginary part if the atom has a net magnetic moment $\vec M$, which in the present case is forced to be parallel to the optical axis of the TEM by the strong magnetic field ($\approx$ 2T) of the objective lens~\cite{RuszPRB07}:  
\begin{equation}
\Im[S({\vec q }, \vec {q'},E)] \propto  (\vec q \times  \vec{q'}) \cdot \vec M \,.
\end{equation}
The dichroic signal is the difference of two spectra - Eq.\ref{DDSCSinter}- obtained by reversing the sign of the third (interference) term. In the difference spectrum the first two terms cancel; only the third one remains. 

%The phase shift is changed by placing the detector in the two positions in the diffraction plane (Fig. 1C) thus selecting two scattering vectors $q \perp q'$ of equal length. Shifting the detector from the "+" to the "-" position in Fig. 1B changes the sign (but not the magnitude) of the vector product $\vec q \times \vec q'$, and thus of the interference term in Eq. \ref{DDSCSinter}. 

The equivalence with XMCD can be understood when considering that the inelastic interaction with a given target atom is  Coulombic. The perturbation, leading to an electronic  transition for an atom at the origin, is an electric field ${\bf F} \propto {\bf q} \, e^{i \omega t + \phi}$ (and similarly for $\bf q'$), with $\hbar \omega=E$, the energy lost by the probe electron in the transition. By forcing the two coherent plane waves (0 and g in Fig. 1) to exhibit a phase difference $\delta\phi=\phi - \phi'=\pi/2$, the electric perturbation vector ${\vec F}+{\vec F}'$ at the  atomic site rotates clockwise in a plane with surface normal $\vec q \times \vec q'$, thereby forcing a chiral transition obeying the selection rule $\Delta m=+1$ (equivalent to the absorption of a photon with positive helicity). When shifting the detector from position "+" to "-" in Fig.~1C, the vector $\vec q \times \vec q'$ changes sign, the perturbation field rotates counterclockwise, and the chirality of the transition is reversed. 
%As with XMCD, the measured difference spectrum is the dichroic signal. 

This ideal situation is never met in practice because it assumes a point-like detector in the diffraction plane measuring a signal from two monochromatic plane waves ionizing an atom at the origin, without any other interaction. However dynamical electron diffraction from the crystal lattice causes a variation of the phase difference $\delta \phi$  along the electron trajectory with a periodicity given by the extinction distance~\cite{RuszPRB07} (which appears as a beating effect in the intensity known as {\em Pendell\"osung} in electron microscopy). Therefore, even using two nearly monochromatic plane waves, the EMCD signal would always be reduced with respect to XMCD, and could even disappear for particular values of the sample thickness. Integration over convergence and collection angles in the microscope/detector system induces additional variations in the phase shift between the coherent partial waves; and the presence of secondary Bragg spots other than 0 and g, contributes to a further  reduction of the dichroic signal.

Diffraction on the crystal lattice, at first view detrimental to the dichroic signal, can be turned to advantage when one realizes that the phase shift between the 0 and the g wave can be tuned by varying the excitation error. Moreover, the lattice periodicity automatically serves as a phase-lock amplifier, creating equal phase shifts in each elementary cell.

When we extend this formalism to the realistic case of many Bragg scattered waves, Eq.~\ref{DDSCSinter} is  replaced by 
\begin{equation}
\frac{\partial^2\sigma}{\partial E\partial\Omega}
=
\frac{4 \gamma^2}{a_0^2 }  \frac{k_f}{k_0}
\sum_{i\leq j}2 \Re[
A_{ij} \frac{S({\vec q_i }, \vec 
{q_j},E)}{q_i^2q_j^2}].
\label{DDSCSinterc}
\end{equation}
where the scattering vectors are enumerated according to the Bragg scattered plane waves in the elastic diffraction pattern. The coefficients $A_{ij}$ are calculated in the framework of dynamical electron diffraction theory~\cite{RuszPRB07b}.

It was found that the dichroic signal is rather robust with respect to variations in incident and detection angle, with only the prefactor varying in magnitude~\cite{LACDIF,RuszPRB07}.
It was therefore tempting to replace the LACDIF~\cite{Morniroli} by a convergent beam diffraction geometry (Fig.1C).
%: here, the specimen remains in eucentric position, and the signal is taken in the diffraction plane. 
Differently from previously reported geometries~\cite{Nature2006,LACDIF} the crystal is tilted to a three-beam case (\textit{i.e.} exciting equally the +g and -g beams).
%, which has a symmetry plane passing through the 0 beam and is perpendicular to $\vec{g}$. 
The diffraction pattern, consisting now of broad Bragg disks instead of sharp point-like spots, is then electronically shifted such that the detector is placed at symmetric positions, labeled "-" and "+" relative to a line perpendicular to the $\bf g$ vector and passing through (000). As opposed to the two-beam case this geometry has the advantage that one avoids any spectral difference not related to dichroism. The EMCD signal has a spatial resolution given by the beam diameter which cannot be reduced below a certain limit because spectra obtained with smaller electron probes have a lower spectral intensity and low SNR. Experimentally it was found that a nominal spot size of 1.7 nm yielded a signal strong enough for detection of EMCD in Fe. 

In order to reliably determine the spatial resolution, a test specimen was produced by means of molecular beam epitaxy. First a 0.8~nm Fe thin film was grown on a (001)-GaAs substrate followed by 25~nm of Au. Then successive Fe and Au layers were stacked as following: 3~nm Fe, 5~nm Au, 6~nm Fe, 10~nm Au, 10~nm Fe, 21~nm Au, 31~nm Fe covered with a 25~nm Au protection layer.

The sample was then prepared in cross section by mechanical grinding and ion-polishing. In order to avoid contamination of the sample in the 200 keV electron beam, the sample was plasma cleaned in 5N Ar atmosphere directly before inserting it into the microscope.  No oxygen was detected during the EELS characterization.

The specimen was first oriented in zone axis conditions, with the Au/Fe interfaces projecting in the TEM image. Then a symmetric 3-beam case was set up, tilting the specimen by roughly 5~degrees off the zone axis such that the interfaces were still projecting, resulting in the excitation of the ($\pm$200) spots. The specimen was characterized with Z-contrast imaging and high-resolution TEM shown in Fig. 1C and Fig. 2A. 

\begin{figure}
\includegraphics[width=\textwidth]{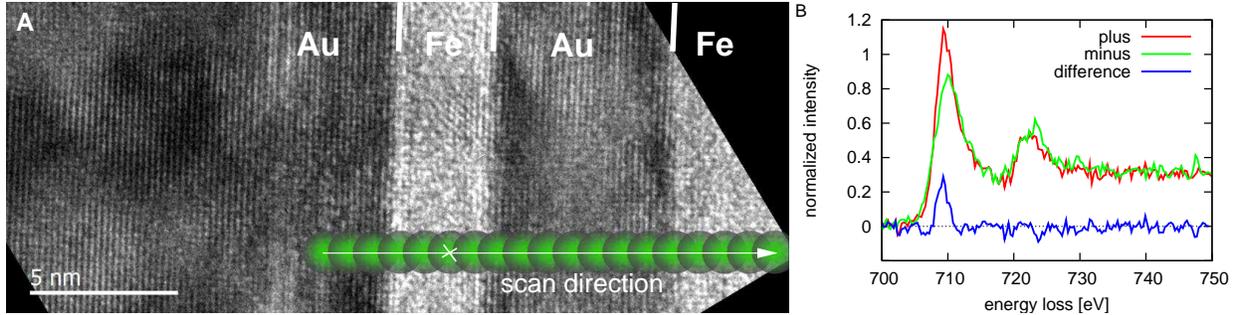}
\caption{\textbf{A:} High-resolution TEM image of the investigated area; the shape and position of the beam during the scan are indicated by the superimposed circles. \textbf{B:} Spectra from the middle of the 3~nm Fe layer, taken at the position marked with a cross. The difference is the dichroic signal.}
\label{fig2}
\end{figure}

The EMCD measurements were performed in the scanning mode of the TEM using the focused probe for two subsequent line scans of the same region, one for each detector position ("+" and "-"). Each line scan, consisting of 20 spectra with a nominal separation of 0.9~nm for a total of 17~nm scan length, started in the first 25~nm Au layer and proceeded on a straight line perpendicular to the Au/Fe interfaces across the first 3~nm Fe layer, the 5~nm Au, the 6~nm Fe, and ended close to the interface with the 10~nm Au layer (Fig. 2A). At each point an energy-loss spectrum at the Fe L$_{2,3}$-edge was acquired, with 10 seconds acquisition time.
These values were chosen to have the highest signal intensity allowed by the specimen and beam drift. The "+" and "-" spectra for the point in the middle of the first Fe layer are shown with their difference, representing the dichroic signal (Fig. 2B).

In Fig.~3 the spectral intensity (\textbf{A}) and the dichroic signal (\textbf{B}), integrated over the L$_3$ edge from 707.9 to 713.9~eV, are shown. The Fe and Au layers are clearly resolved, thus demonstrating a spatial resolution of at least 3~nm.

\begin{figure}
\includegraphics[width=1 \textwidth]{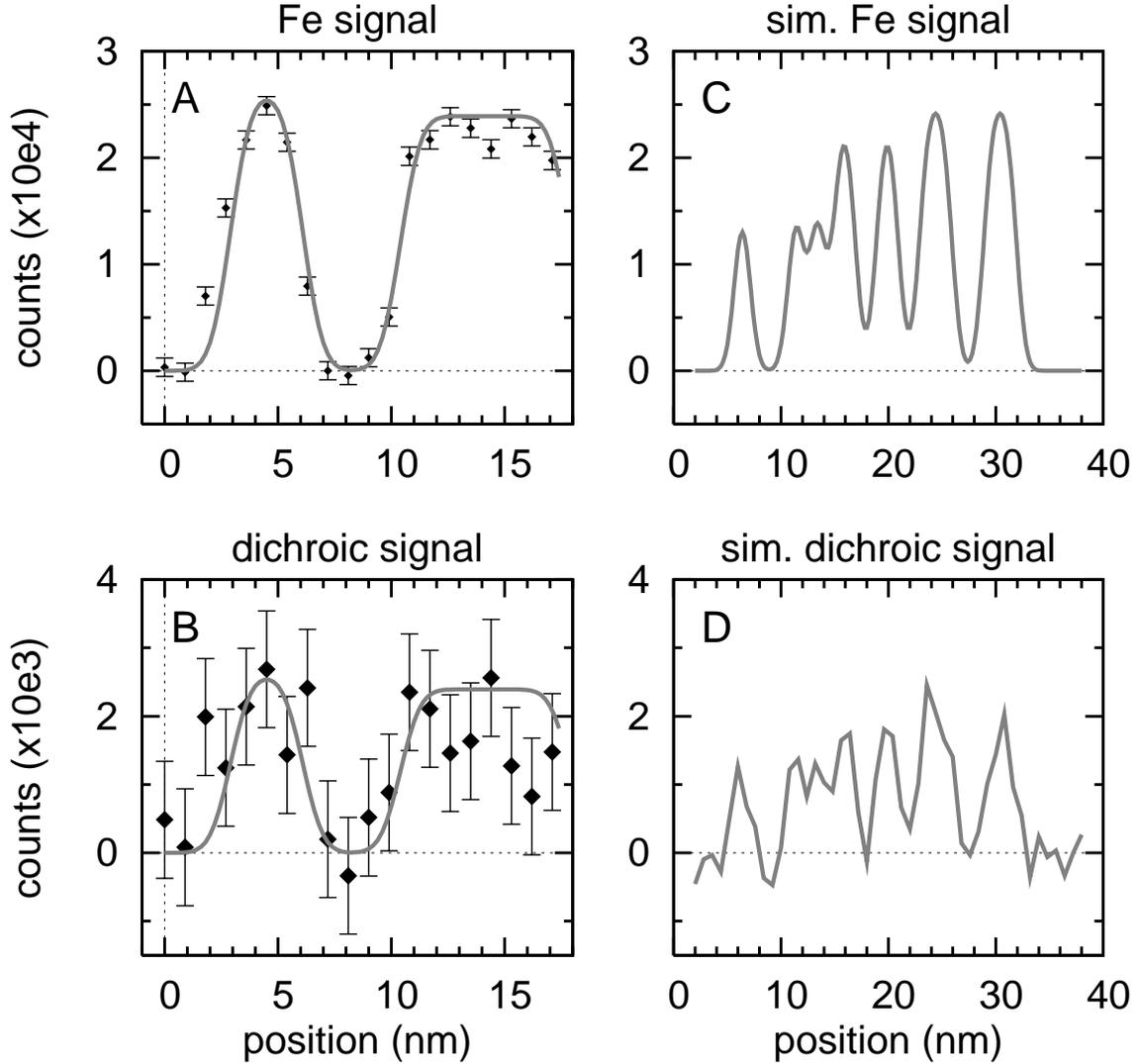}
\caption{\textbf{A:} Profile of the line scanned in Fig. 2. The experimental points are the integrated Fe signal (sum of the "+" and "-" spectra) at the L$_3$ edge (707.9-713.9~eV). The best fit with a Gaussian spot shape gives a FWHM of 1.66~nm for the spot. The error bars are 3$\sigma$~=~855~counts. \textbf{B:} Corresponding line profile of the dichroic signal (difference of the "+" and "-" spectra) integrated at the L$_3$ edge. \textbf{C, D:} Simulated Fe (C) and EMCD (D) profiles for a stack of 4~nm Au,1~nm Fe, 4~nm Au, followed by double layers of 1~nm Fe/Au, 2~nm Fe/Au and 3~nm Fe/Au. The EMCD profile is shown with simulated Poissonian noise corresponding to the present experimental conditions, demonstrating a resolution of $\approx$ 2~nm.}
\label{fig3}
\end{figure}

For the determination of the effective resolution of the EMCD experiment a Gaussian spot profile  sweeping across the Au/Fe multilayer is assumed. A least-squares fit to the experimental values yields a variance of $\sigma^2=1.0$~nm$^2$ which translates into a Gaussian full width at half maximum (FWHM) of 1.66~nm. This proves that the factor limiting the resolution is indeed the spot size; delocalization\footnote{The delocalization via the long range Coulomb interaction is given by the modified Bessel function $K_0(q_E x)$ where $q_E=k_0 E/2E_0$ is the characteristic wave number, \textit{i.e.} for the Fe L edge and 200keV primary electron energy $q_E=4.4/nm$. The FWHM of the delocalization function is 0.08~nm, and the 10 \% level is at 0.6 nm~\cite{SchattPRB99}} or non-projecting interfaces are negligible in the present case. The deviations from the fit function in the leftmost slope in the figure are caused by inconstant drift of the specimen during the scan;  variations on the plateau to the right stem from faint thickness variation resulting in changes of the peak height. 

We conclude that the Fe signal can be detected with a resolution limit of  $< 1.7$~nm in the present experiment. Since the EMCD is a difference of Fe signals, its theoretical geometric resolution must be the same. But this is only true for the same SNR. The smaller SNR in the EMCD signal reduces this limit. Nonetheless, the EMCD signal is clearly visible across the 3~nm Fe layer in Fig.~3B.

%To calculate the resolution limit for EMCD we would need a test sample with several alternating thin layers (ideally starting from 1~nm and then slowly increasing; the spatial resolution is then given by the thinnest layer that can be resolved).In the present case we refer to estimates based on the experiment. 

The inherent spatial resolution of the experiment derives from  the variance of the signal along the scan, invoking the Rose criterion~\cite{Roseb} for detection of a faint object in a noisy data set. 
%It states that the signal $I$ defining the object  must exceed the background noise by a factor of at least 3 in order to be recognized as a structure. 
In doing so, we first simulated the Fe L$_3$ profile of a hypothetical multilayer with varying layer thickness for the  actual spot diameter, as shown in Fig.~3C.
%starting with 4~nm Au, followed by 1~nm Fe, 4~nm Au, 1~nm Fe, 1~nm Au, 1~nm Fe, 1~nm Au, 2~nm Fe, 2~nm Au, 2~nm Fe, 2~nm Au, 3~nm Fe, 3~nm Au, 3~nm Fe (from left to right). 
Panel D shows the corresponding simulated noisy EMCD data set for the given experimental conditions. The single free-standing layer of 1 nm Fe can be detected clearly, but a stack of 1 nm Fe layers separated by 1 nm Au gives an unstructured EMCD signal. Individual layers are visible with at least 2 nm separation.

We conclude that with the present method EMCD signals can be detected with a spatial resolution of $2$~nm. This constitutes  a breakthrough for the study of nanomagnetism at interfaces and boundaries. 
The main limiting factors in this experiment are specimen drift and beam instability, which set upper bounds on the collection time of the spectra. If drift can be reduced the dwell time could be increased, thereby lowering the noise level and  allowing  smaller spots. With the new generation of $C_s$ corrected (scanning-)TEMs it is likely to achieve sub-nm resolution in EMCD spectrometry. 

%\bibliography{chiralstem}

\begin{acknowledgments}
The authors acknowledge N. J. Zaluzec, C. H{\'e}bert and J. Verbeeck for programming the script for the electronic shift of the diffraction pattern; and E. Carlino, L. Felisari, F. Maccherozzi and P. Fischer for fruitful discussions. P.S. and S.R. acknowledge funding from the European Union under contract nr. 508971 (FP6-2003-NEST-A) "CHIRALTEM".
\end{acknowledgments} 

\end{document}